\documentclass[twocolumn,aps,floatfix,superscriptaddress]{revtex4}
\pdfoutput=1 
\usepackage{amsmath,amssymb,eucal,graphicx,float,epstopdf}

\usepackage{hyperref}
\usepackage{subfigure}
\usepackage{color}
\usepackage{float}
\usepackage{url}

\begin{document}
\title{Exclusion in Junction Geometries}
\author{Keming Zhang}
\affiliation{Department of Computer Science, Rice University, Houston,  TX, 77005
  USA}
\author{P. L. Krapivsky}
\affiliation{Department of Physics, Boston University, Boston, MA, 02215 USA}
\author{S. Redner}
\affiliation{Santa Fe Institute, 1399 Hyde Park Road, Santa Fe, NM, 87501 USA}

\begin{abstract}

  We investigate the dynamics of the asymmetric exclusion process at a
  junction.  When two input roads are initially fully occupied and a single
  output road is initially empty, the ensuing rarefaction wave has a rich
  spatial structure.  The density profile also changes dramatically as the
  initial densities are varied.  Related phenomenology arises when one road
  feeds into two.  Finally, we determine the phase diagram of the open
  system, where particles are fed into two roads at rate $\alpha$ for each
  road, the two roads merge into one, and particles are extracted from the
  single output road at rate $\beta$.

\end{abstract}

\maketitle

\section{Introduction}
\label{sec:intro}

In exclusion processes, sites can be occupied by at most one particle and
particles hop to empty sites.  This paradigmatic model sheds much light on
non-equilibrium steady states, large deviations, and other aspects of
strongly-interacting infinite-particle systems (see, e.g.,
\cite{Spohn91,SZ95,D98,S00,BE07,D07,book} and references therein).  The
totally asymmetric simple exclusion process (TASEP), where particles can hop
to neighboring empty sites in one direction, is a minimalist realization of
exclusion processes that is particularly tractable and also has a diverse
range of applications~\cite{MGP68,CSS00,CMZ11,AES15}.

\begin{figure}[ht]
\centerline{\subfigure[]{\includegraphics[width=0.23\textwidth]{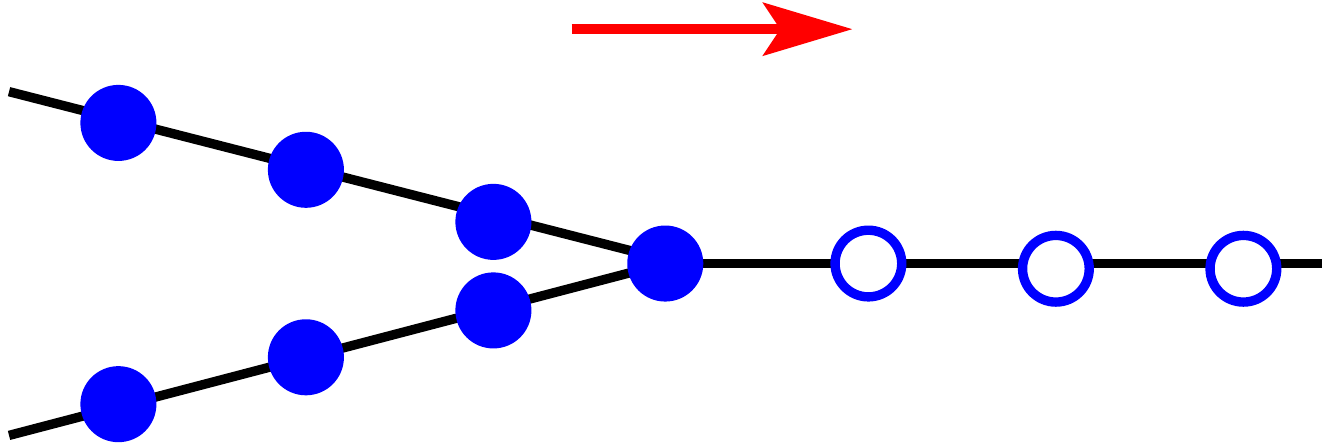}}\quad
\subfigure[]{\includegraphics[width=0.23\textwidth]{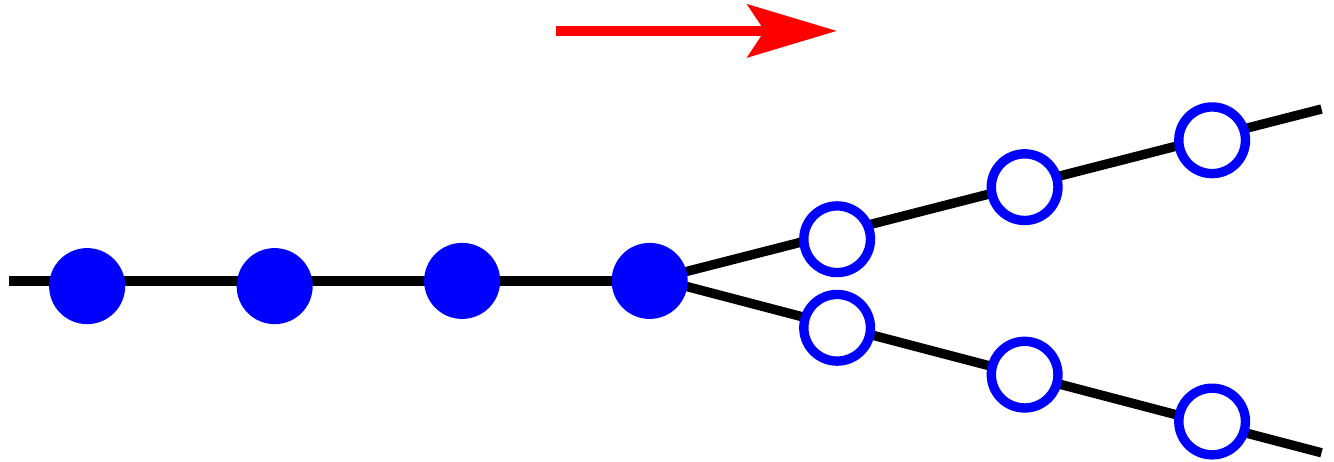}}}
\caption{Illustration of the TASEP at: (a) a $(2,1)$ junction and (b) a
  $(1,2)$ junction.  Shown is the downstep initial condition in which sites
  are fully occupied for $x<0$ (solid circles) and empty otherwise (open
  circles).}
  \label{fig:junction}
\end{figure}

In this work, we investigate the properties of the TASEP at a
\emph{junction}, where a small number of incoming roads, that each carry a
TASEP, meet at a single point and particles leave via an outgoing road (or
roads) also by the TASEP (Fig.~\ref{fig:junction}).  Our initial motivation
came from the observation of maddening delays that arise when disembarking
from a passenger plane.  Here, the aisle(s) get clogged with passengers who
are either slow in retrieving their belongings or in walking, leading to a
clogging at the exit door of the plane.  Our junction TASEP model is a rough
caricature for this disembarkment process.  We study in detail
(Sec.~\ref{sec:(2,1)-RW}) the $(2,1)$ junction geometry with two roads that
start at $x=-\infty$ and merge at $x=0$ into a single outgoing road that
extends to $x=+\infty$.  We also analyze the $(1,2)$ junction geometry
(Sec.~\ref{sec:(1,2)-RW}) and finite systems (Sec.~\ref{sec:(2,1)-open}).
Our analysis can be generalized to other junction geometries.

For the $(2,1)$ junction geometry, one might expect a pileup of particles as
the junction is approached, reminiscent of what occurs when highway traffic
approaches a lane constriction.  The role of blockage in the TASEP has been
considered previously in a one-dimensional geometry in which the hopping rate
of a single bond is reduced from 1 to $r<1$~\cite{JL92,JL94,SPS15}.  For the
slow bond problem, particles jam upstream from the blockage, as one might
anticipate, and the focus of~\cite{JL92,JL94,SPS15} was to characterize this
jam (see also Refs.~\cite{K00,B06,CLST13} for reviews of this problem).  A
complementary TASEP model, in which particles may fly to any empty site when
they reach a single special site was introduced in~\cite{ABKM13}.  All these
studies focused on stationary properties.  The slow bond problem with the
domain wall initial condition was studied in~\cite{L99,BSS14,BSS17}.  A
related line of research has focused on the role of an intersection on the
TASEP, with the same number of incoming and outgoing roads at a
junction~\cite{N93,IF96,FSS04,FN07,BF08,EPK09,JA12,RPK13};
Refs.~\cite{EPK09,RPK13} study aspects of the TASEP in the junction
geometries similar to those considered in our work.

In what follows, we assume that the hopping rates at each site are the same
in all roads, and set them equal to 1.  Thus each particle can hop to the
right only if its right neighbor is empty.  We define the location of the
junction as the origin.  We employ a hydrodynamic description and use the
continuous density $\rho(x,t)$ as the basic dynamical variable in the
long-time limit.

For the ``downstep'' initial condition in the $(2,1)$ junction geometry, in
which each site on the two incoming roads are initially occupied while the
single outgoing road is empty, the density profile at long times contains
both a constant-density jammed segment upstream from the junction, as well as
a downstream linear rarefaction wave (Fig.~\ref{rho21-1}).  As the initial
density in the incoming roads is decreased, the form of the rarefaction wave
changes dramatically and a shock wave can even arise.  Similarly rich
phenomenology arises for the $(1,2)$ junction geometry.  Finally, we study
the open $(2,1)$ system in which current is fed in to the system at rate
$\alpha$ at each upstream road far from the junction and current is extracted
at rate $\beta$ in the single road far downstream from the junction.  We map
out the phase diagram of this system and highlight the differences with the
open TASEP system on the line.

\section{Shock and Rarefaction Waves}
\label{sec:S-RW}

As a preliminary, we recapitulate the well-known (see, e.g., \cite{book})
density profile that arises in the TASEP on the line for the initial density
step
\begin{equation}
\label{step}
\rho = 
\begin{cases}
\rho_{\scriptstyle L} & x<0\,,\\[1mm]
\rho_{\scriptstyle R} & x>0\,,
\end{cases}
\end{equation}
where $\rho_{\scriptstyle L}$ and $\rho_{\scriptstyle R}$ are constant
densities to the left and to the right of the step.  The hydrodynamic
behavior is governed by the continuity equation
\begin{equation}
\label{CE}
\frac{\partial\rho}{\partial t} +\frac{\partial j}{\partial  x} =0\,,
\end{equation}
with the current given by $j=\rho(1-\rho)$.  The solution to Eq.~\eqref{CE} 
subject to \eqref{step} has a remarkably simple scaling form 
\begin{equation}
\label{scaling}
\rho(x,t) = f(z), \quad z=x/t\,.
\end{equation}
When this scaling form is substituted into \eqref{CE}, two distinct behaviors
arise that depend on whether $\rho_{\scriptstyle L}>\rho_{\scriptstyle R}$ or
$\rho_{\scriptstyle L}<\rho_{\scriptstyle R}$:
\begin{itemize}

\item {\bf Rarefaction wave} ($\rho_{\scriptstyle L}>\rho_{\scriptstyle R}$).
  An initial downstep relaxes to the rarefaction wave
\begin{subequations}
    \begin{equation}
\label{RW}
\rho(x,t) = 
\begin{cases}
\rho_{\scriptstyle L}      & z<1\!-\!2\rho_{\scriptstyle L}\,,\\[1.5mm]
\frac{1}{2}(1-z)   &1\!-\!2\rho_{\scriptstyle L}< z < 1\!-\!2\rho_{\scriptstyle R}\,,\\[1.5mm]
\rho_{\scriptstyle R}    & z>1\!-\!2\rho_{\scriptstyle R}\,.
\end{cases}
\end{equation}

\item {\bf Shock wave} ($\rho_{\scriptstyle L}<\rho_{\scriptstyle R}$).  An
  upstep persists as a shock wave and merely translates:

\begin{equation}
\label{SW}
\rho(x,t) = 
\begin{cases}
\rho_{\scriptstyle L} & z<c\,,\\[1mm]
\rho_{\scriptstyle R} & z>c\,,
\end{cases}
\end{equation}
with shock speed $c = 1- \rho_{\scriptstyle R} - \rho_{\scriptstyle L}$.
\end{subequations}
\end{itemize}

\section{Rarefaction at a (2,1) Junction}
\label{sec:(2,1)-RW}

We now investigate the evolution of the initial density step \eqref{step} at
a junction where two roads merge into one.  Here, and in the following
section, the system is unbounded, with the incoming road(s) extending to
$x=-\infty$ and the outgoing road(s) extending to $x=+\infty$.   For simplicity,
we treat the special case where the outgoing road is empty,
$\rho_{\scriptstyle R}=0$, and where the initial densities in the incoming
roads are both equal to $\rho_{\scriptstyle L}$.  Three distinct behaviors
arise for: (a) $\rho_{\scriptstyle L}>\rho_+$, (b)
$\rho_+>\rho_{\scriptstyle L}>\rho_-$, and (c)
$\rho_{\scriptstyle L}<\rho_-$, where $\rho_+$ and $\rho_-$ are critical
densities whose values are given in Eq.~\eqref{rho-crit} below.  We discuss
these three cases in turn.

\subsection{High input density: $\rho_{\scriptstyle L} \geq \rho_+$}

\begin{figure}[ht]
  \subfigure[]{\includegraphics[width=0.375\textwidth]{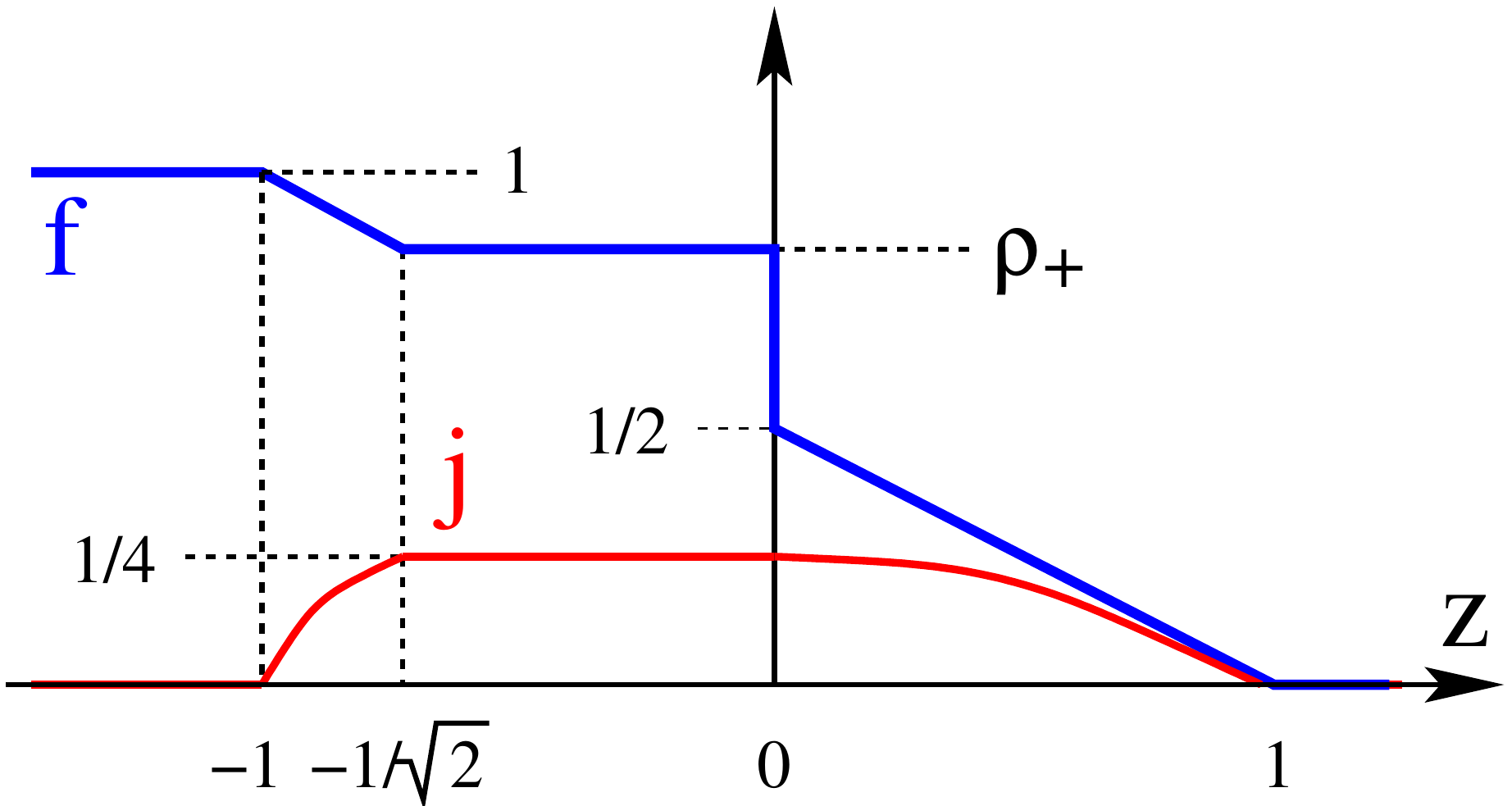}}\qquad\qquad
    \subfigure[]{\includegraphics[width=0.375\textwidth,height=3.75cm]{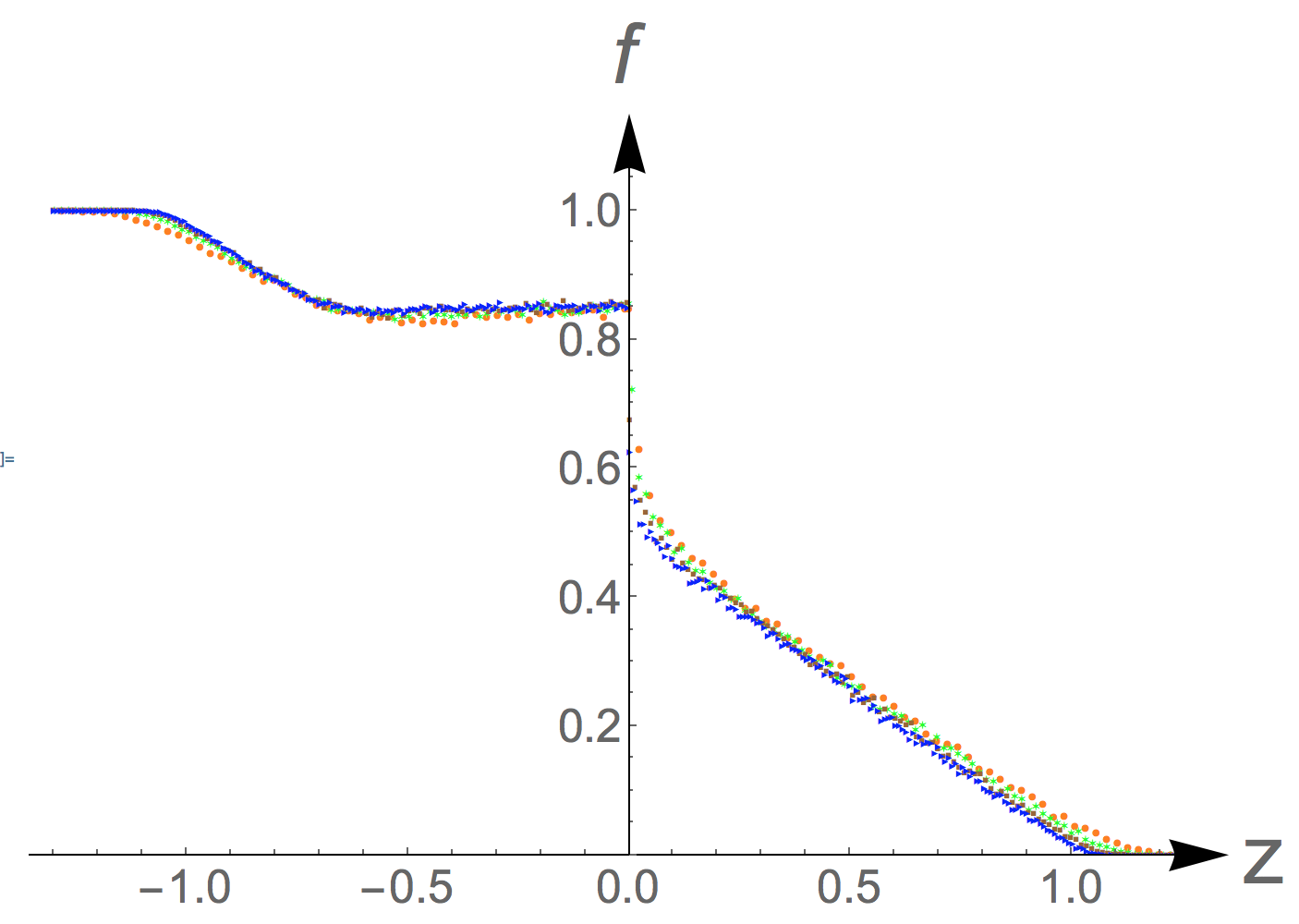}}
    \caption{ (a) Schematic, but to scale, scaled density profile for a
      rarefaction wave (blue) at a $(2,1)$ junction for an initial density
      downstep with $\rho_{\scriptstyle L}=1$.  The corresponding current is
      shown in red.  (b) Simulation data for the scaled density profile for
      this same initial condition for $t=125$ (red), $250$, (green) $500$
      (brown), and $1000$ (blue). }
  \label{rho21-1}
\end{figure}

As in the conventional TASEP, a density downstep develops into a rarefaction
wave in the subrange $0<x<t$.  However, for
$\rho_{\scriptstyle L} \geq \rho_+$, a density pileup develops just upstream
from the junction, with a sudden density drop at $x=0$ (Fig.~\ref{rho21-1}).
This same qualitative behavior occurs as long as the initial input density
$\rho_{\scriptstyle L}$ is greater than $\rho_+$.  Upstream from the pileup,
the density profile is again given the the classic rarefaction wave.

This rich behavior can be readily understood in the hydrodynamic limit.
Substituting the scaling form \eqref{scaling} into the continuity
equation~\eqref{CE} shows that the scaling function satisfies
\begin{align}
  \label{fp}
  f'(z)\big[1-2f(z)-z\big]=0\,,
\end{align}
where the prime denotes differentiation with
respect to $z$.  The only solutions to this equation are either:
\begin{align*}
  &\text{flat\ profile}\quad\qquad f'(z)=0,\\
   &\text{linear\ profile}\quad\quad  f=\tfrac{1}{2}(1-z).
\end{align*}

Any solution is a combination of these elemental forms.  For the conventional
TASEP, the solutions are the aforementioned rarefaction and shock wave
solutions, Eqs.~\eqref{RW}--\eqref{SW}.  To determine the rarefaction wave at
a $(2,1)$ junction, note that the current just to the right of the junction
cannot exceed the maximum possible value of $j_{\rm max}=\frac{1}{4}$.  If
one starts at $z=-1$ (equivalently $x=-t$), and increases $z$, the density
decays from +1 as $\rho(z)=\frac{1}{2}(1-z)$.  Correspondingly, the current
increases as $j(z)=2\rho(1-\rho)= 2\times\frac{1}{4}(1-z^2)$, where the
prefactor 2 accounts for the two upstream roads.  Because the current cannot
exceed $\frac{1}{4}$, $j(z)$ must ``stick'' at this value when first reached,
which happens when $z=-1/\sqrt{2}$.  Thus for $z$ in the range
$-1/\sqrt{2}<z<0$, the density also sticks at a pileup value that corresponds
to this maximal current.  This maximal current condition is
$2\rho(1-\rho)=\frac{1}{4}$, with solutions
\begin{equation}
\label{rho-crit}
\rho_+=\frac{1}{2}+\frac{1}{\sqrt{8}}\,, \qquad \rho_-=\frac{1}{2}-\frac{1}{\sqrt{8}}\,.
\end{equation}
It is the larger root $\rho_+\approx 0.853553$ that is realized for the
downstep initial condition, as shown in Fig.~\ref{rho21-1}.

For $z=0^+$, there is only a single road and the density must suddenly drop
to $\frac{1}{2}$, so that the current at $z=0^+$ matches the maximal current
$j_{\rm max}=\frac{1}{4}$ at $z=0^-$.  For $0<z<1$, the density decays
linearly with $z$ until the density reaches 0 at $z=1$. Thus in the
high-density regime defined by $\rho_{\scriptstyle L} \geq \rho_+$, we
conclude that the scaled density profile consists of five distinct segments:
\begin{equation}
\label{RW:1}
f = 
\begin{cases}
\rho_{\scriptstyle L}        & \quad z<1-2\rho_{\scriptstyle L}\\[1mm]
\frac{1}{2}(1-z)       & \quad1-2\rho_{\scriptstyle L} < z < - \frac{1}{\sqrt{2}}\\[1mm]
\rho_+        & \quad -\frac{1}{\sqrt{2}} < z < 0\\[1mm]
\frac{1}{2}(1-z)       & \quad 0<z<1 \\[1mm]
0             &\quad z>1.
\end{cases}
\end{equation}
Simulation data converge to this five-segment form, with finite-time
corrections that systematically vanish as $t$ increases
(Fig.~\ref{rho21-1}(b)).  For the step initial condition, the system length is
effectively infinite because the spatial range over which the density is
varying is less than the actual system length. 

\subsection{Intermediate input density: $\rho_-\leq \rho_{\scriptstyle L} \leq \rho_+$}

\begin{figure}[ht]
  \subfigure[]{\includegraphics[width=0.375\textwidth]{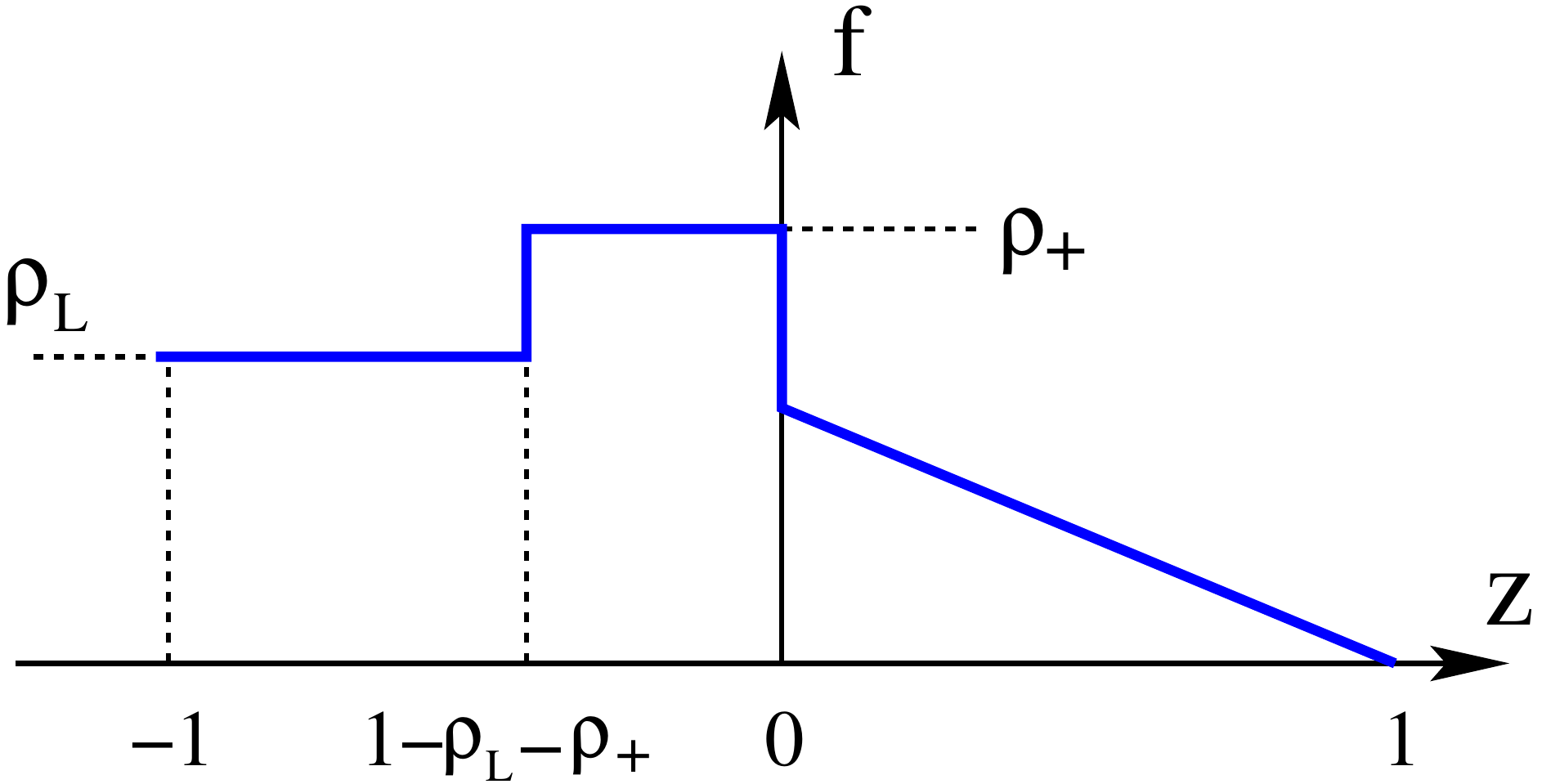}}\qquad\qquad
    \subfigure[]{\includegraphics[width=0.375\textwidth,height=3.75cm]{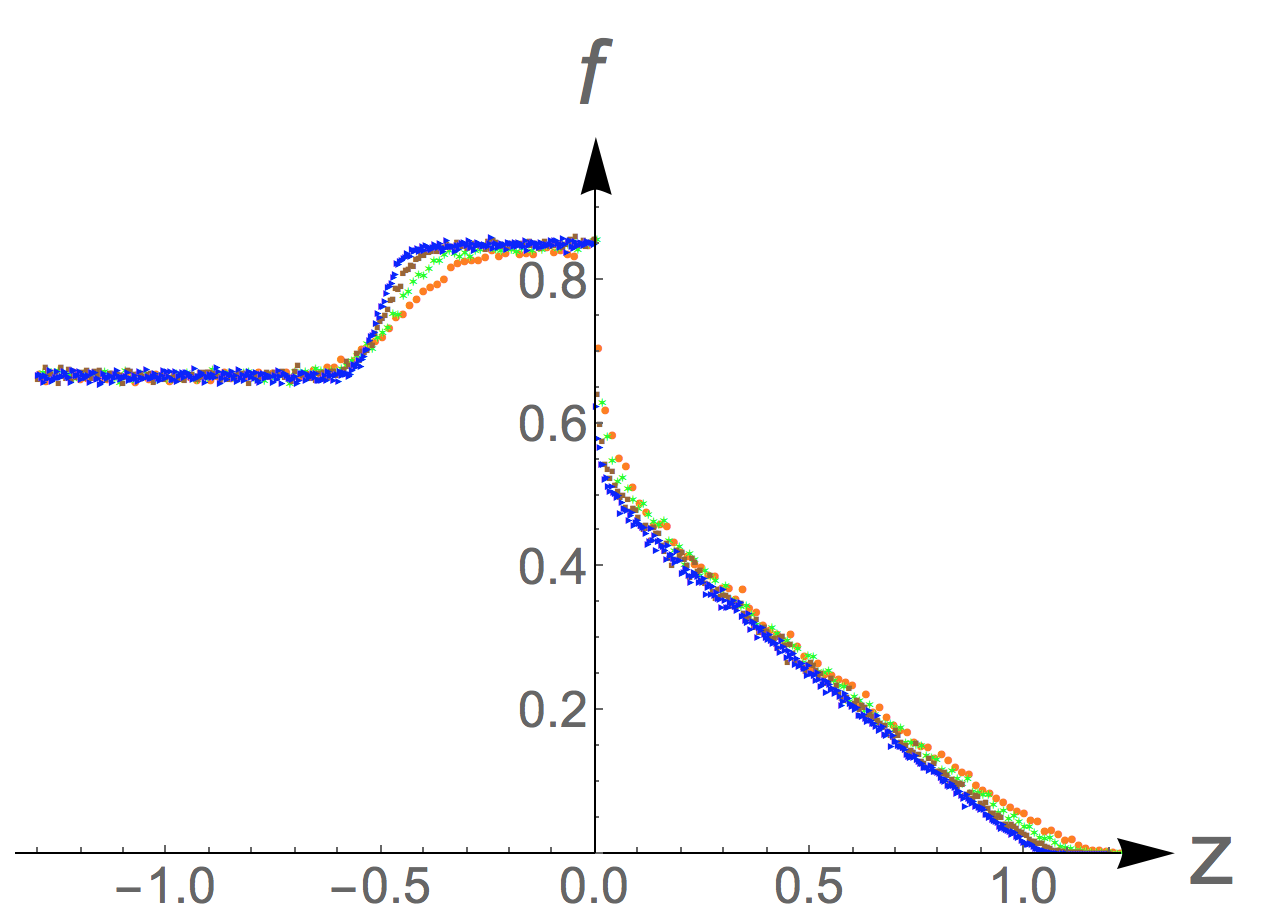}}
    \caption{(a) The scaled density profile in Eq.~\eqref{RW:2} for the
      $(2,1)$ junction with $\rho_{\scriptstyle L}=2/3$.  (b) Simulation data
      for this same initial condition for $t = 125$ (red dots), $250$
      (green), $500$ (brown), and $1000$ (blue).  The linear rise in the
      density near $z=1-\rho_{\scriptstyle L}-\rho_+$ gradually steepens for
      increasing $t$, showing that this behavior is a finite-size effect. }
  \label{rho23}
\end{figure}

Distinct behavior arises when $\rho_{\scriptstyle L}$ lies between the two
critical values $\rho_+$ and $\rho_-$.  For $\rho_{\scriptstyle L}$ in this
range, the current in each incoming road is less than $j_{\rm max}$, but the
sum of the currents in the two roads exceeds $j_{\rm max}$.  Thus there again
must be a pileup of particles upstream from the junction point, as the
maximum current that can be transmitted at the junction is
$j_{\rm max}=\frac{1}{4}$.  To match the outgoing current at the junction,
the pileup density must equal $\rho_+$.  On the other hand, the asymptotic
density for $z\to -\infty$ is $\rho_{\scriptstyle L}$.  As a result, a shock
wave must arise whose speed is given by $1-\rho_+-\rho_L$. Thus when
$\rho_-< \rho_{\scriptstyle L} <\rho_+$, the asymptotic density profile
consists of four segments:
\begin{equation}
\label{RW:2}
f = 
\begin{cases}
\rho_{\scriptstyle L}       & \quad z<1-\rho_+-\rho_{\scriptstyle L} \\
\rho_+                      &  \quad 1-\rho_+-\rho_{\scriptstyle L}  < z < 0\\
\frac{1}{2}(1-z)             &  \quad 0<z<1 \\
0                           &  \quad z>1\,.
\end{cases}
\end{equation}
Even though this initial density downstep leads to a rarefaction wave in the
classic TASEP, the road constriction leads to a jam the manifests itself as a
left-moving shock wave on the upstream side of the junction.  A typical
example profile for the case of $\rho_{\scriptstyle L}=2/3$ is shown in
Fig.~\ref{rho23}.

\subsection{Low input density: $\rho_{\scriptstyle L} \leq \rho_-$}

Finally, we treat the low input density regime, where the incoming density obeys 
$\rho_{\scriptstyle L}\leq \rho_-$.  Now the total current in the two
incoming roads is always less than or equal to $j_{\rm max}=\frac{1}{4}$.
Consequently, all the incoming current can be accommodated by the single
output road.  Therefore, there is no pileup at $x=0$ and the density profile
in the incoming roads does not change in time.

In the special case of $\rho_{\scriptstyle L}=\rho_-$,
the total input current to the junction equals $j_{\rm max}=\frac{1}{4}$,
corresponding to the maximum current that can be accommodated by the output
road.  Here, the density profile for $z>0$ is again the classic rarefaction
wave.  When $\rho_{\scriptstyle L}<\rho_-$, the input current to the junction,
$2\rho_{\scriptstyle L}(1-\rho_{\scriptstyle L})$, is less than
$j_{\rm max}$.  To have a consistent scaling solution for $z>0$, there must
be a flat profile immediately to the right of the junction, with density
$\rho_{\scriptstyle R}$, that eventually joins to the rarefaction wave
$\rho(z)=\frac{1}{2}(1-z)$.  We determine the density in the flat region to
the right of the junction by matching the input and outgoing currents at
$z=0$.  This yields
\begin{subequations}
\begin{equation}
\label{jlr}
j_{\rm in}= 2\rho_{\scriptstyle L}(1-\rho_{\scriptstyle L})
  = \rho_{\scriptstyle R}(1-\rho_{\scriptstyle R})
\end{equation}
from which 
\begin{align}
  \label{rhorl}
  \rho_{\scriptstyle R} = \frac{1-\sqrt{1-8 \rho_{\scriptstyle L}(1-\rho_{\scriptstyle L})}}{2}\,.
\end{align}
\end{subequations}

\begin{figure}[ht]
    \subfigure[]{\includegraphics[width=0.375\textwidth]{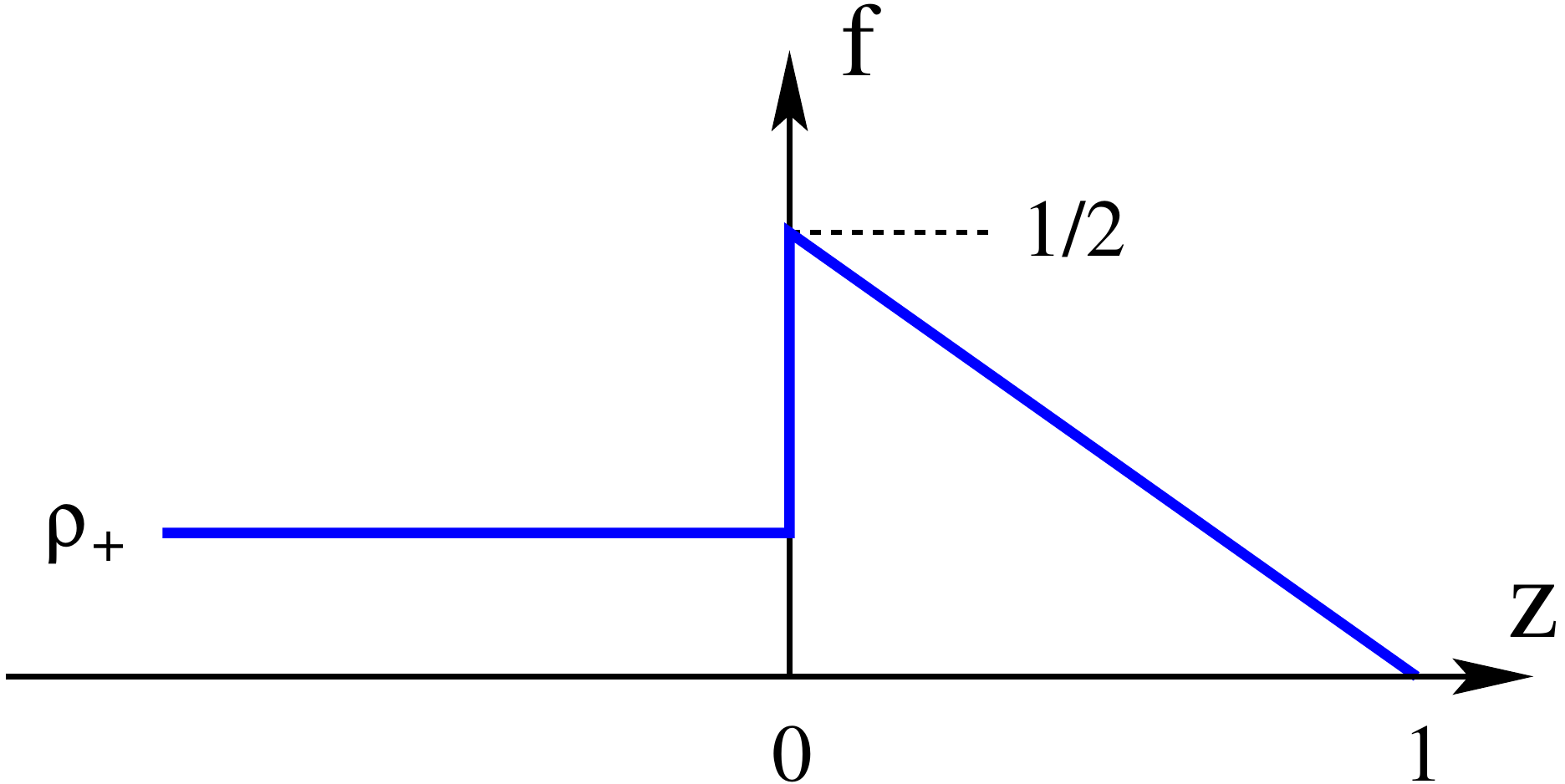}}
    \subfigure[]{\includegraphics[width=0.375\textwidth]{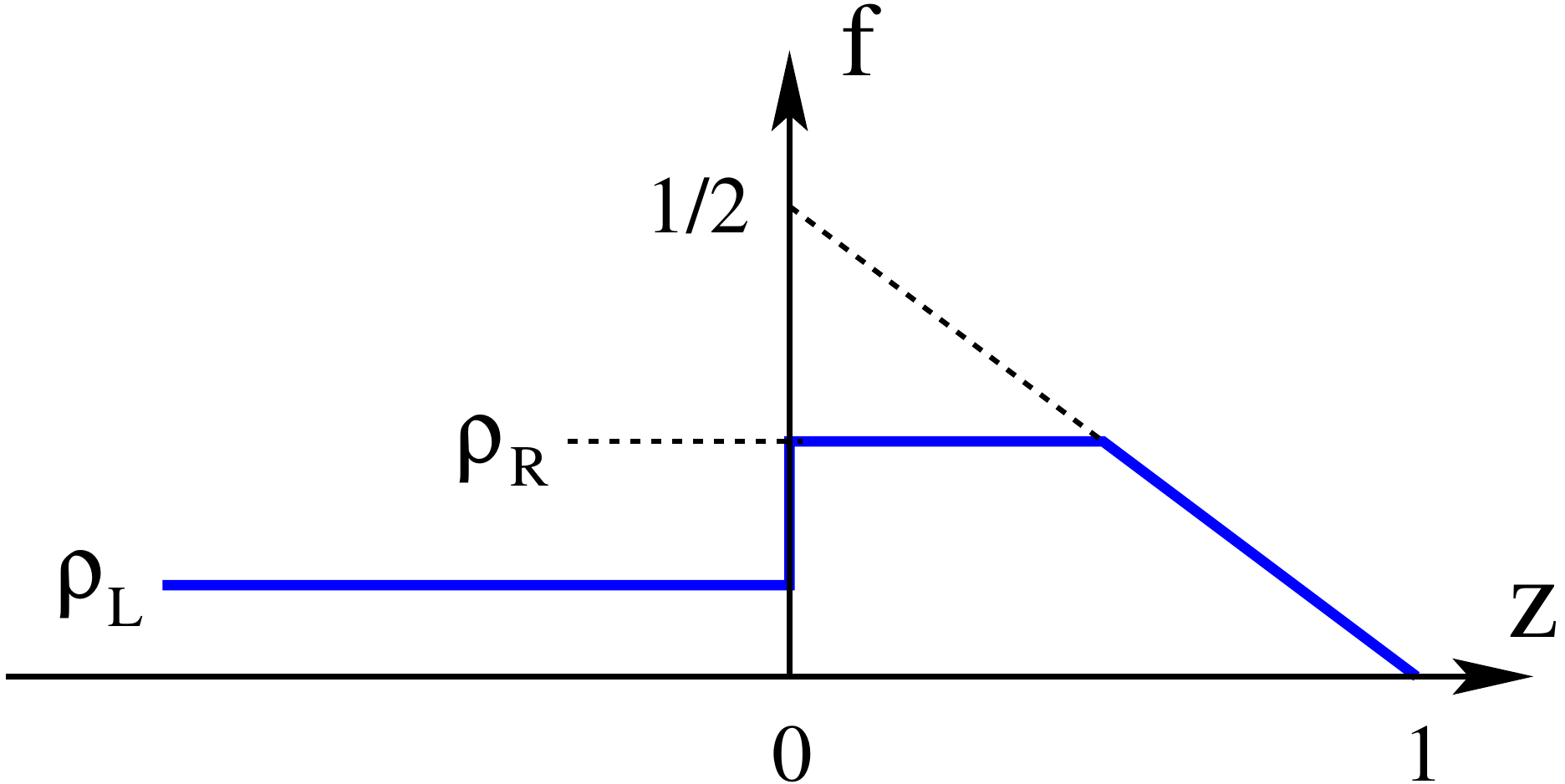}}
    \caption{The scaled density profile for: (a)
      $\rho_{\scriptstyle L}=\rho_-\approx 0.146$ and
      $\rho_{\scriptstyle L}=\frac{1}{10}$.  }
  \label{small-rho}
\end{figure}

Assembling these results, the scaled density profile consists of three segments when
$\rho_{\scriptstyle L}=\rho_+$ (Fig.~\ref{small-rho}(a)):
\begin{subequations}
\begin{equation}
\label{RW:3a}
f = 
\begin{cases}
\rho_{\scriptstyle L}                          & \quad z<0\\
\frac{1}{2}(1-z)                           &\quad 0 <z<1 \\
0                                          &  \quad z>1\,,
\end{cases}
\end{equation}
and four segments for $\rho_{\scriptstyle L}<\rho_+$ (Fig.~\ref{small-rho}(b)):
\begin{equation}
\label{RW:3b}
f = 
\begin{cases}
\rho_{\scriptstyle L}                & \quad z<0\\
\rho_{\scriptstyle R}     & \quad 0 < z < 1-2\rho_{\scriptstyle R}\\
\frac{1}{2}(1-z)               &\quad 1-2\rho_{\scriptstyle R}<z<1 \\
0                                          &  \quad z>1\,.
\end{cases}
\end{equation}
\end{subequations}

\section{Rarefaction at a (1,2) Junction}
\label{sec:(1,2)-RW}

\begin{figure}[ht]
\subfigure[]{\includegraphics[width=0.375\textwidth,height=3.75cm]{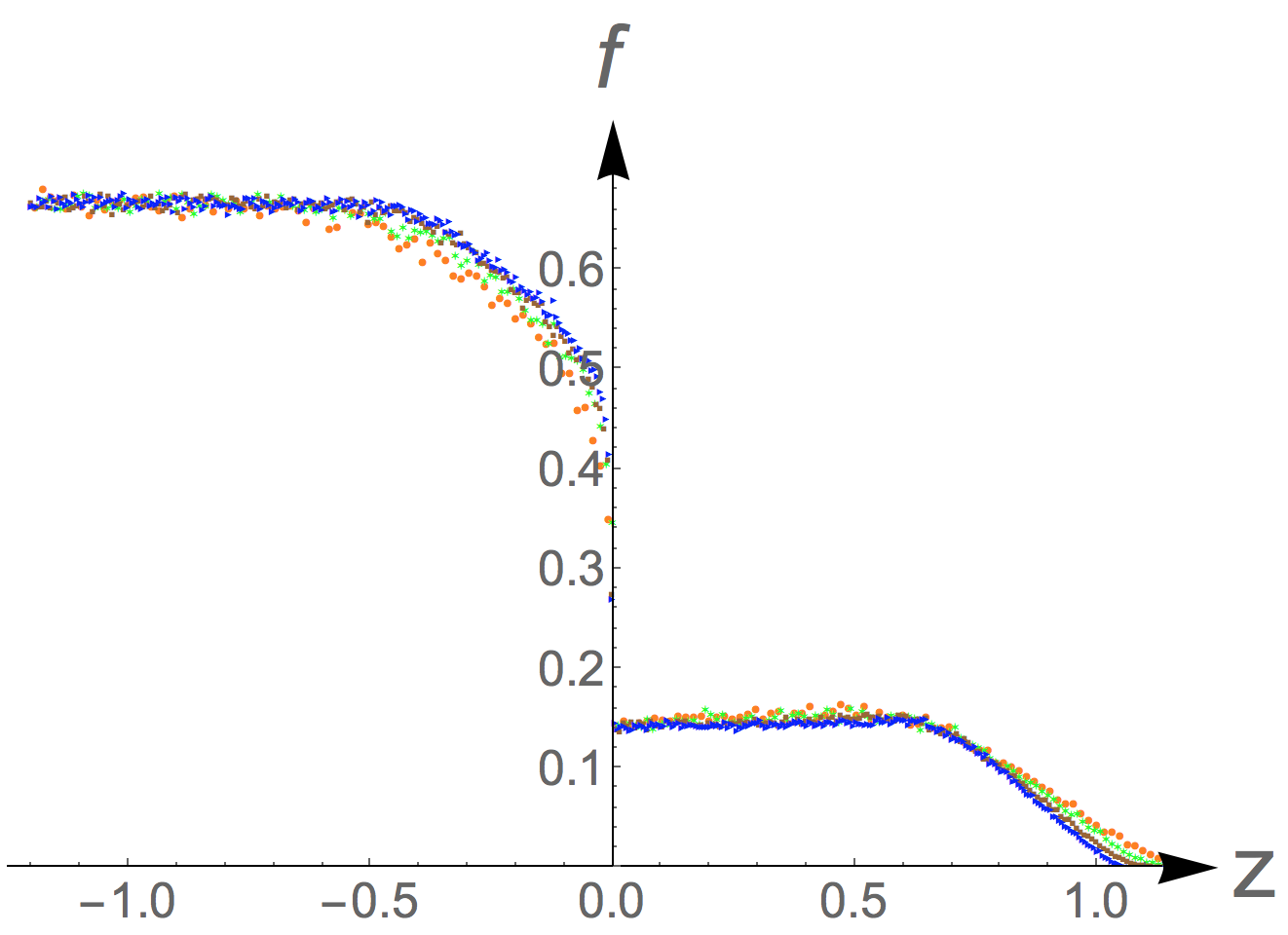}}\qquad\qquad
 \subfigure[]{\includegraphics[width=0.375\textwidth,height=3.75cm]{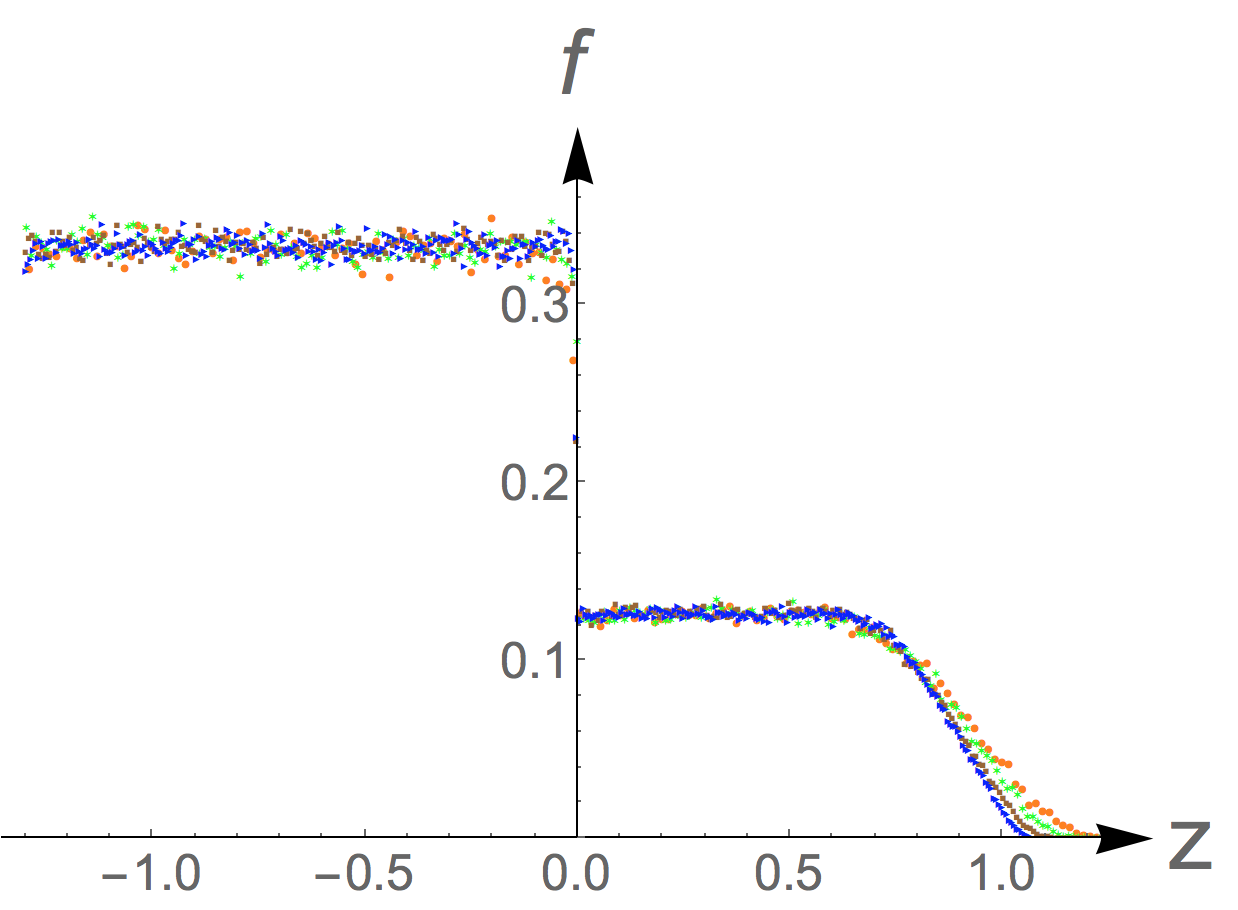}}\qquad\qquad
 \caption{Simulation data for the density profile for the $(1,2)$ junction,
   for (a) $\rho_{\scriptstyle L}=2/3$ and (b) $\rho_{\scriptstyle L}=1/3$
   for $t = 125$ (red dots), 250 (green), 500 (brown), and 1000 (blue).}
  \label{rho-1to2}
\end{figure}

The same type of arguments as those given above can be applied to the $(1,2)$
junction (see Fig.~\ref{fig:junction}(b)).  It is again natural to consider the
initial condition of $\rho=\rho_{\scriptstyle L}$ for $z<0$ and $\rho=0$ for
$z>0$ and study the behavior as a function of $\rho_{\scriptstyle L}$.  As in
the $(2,1)$ junction, a rich set of behaviors arises for varying
$\rho_{\scriptstyle L}$ (Fig.~\ref{rho-1to2}).

For the initial state where $\rho_{\scriptstyle L}=1$, that is, the input
road is fully occupied and the two downstream roads are empty, we can exploit
particle-hole duality of the TASEP to immediately infer the density profile.
In this duality, a particle moving to the right corresponds to a hole (a
vacancy) moving to the left.  The density of holes $\rho_h$ is related to the
particle density $\rho$ by $\rho_h=1-\rho$.  Thus the dynamics of a
right-moving TASEP at a $(1,2)$ junction with the downstep initial state of
$\rho=1$ for $z<0$ and $\rho=0$ for $z>0$ is equivalent to the TASEP dynamics
of holes that move to the left in the $(2,1)$ junction geometry with $\rho_h=1$
for $z>0$ and $\rho_h=0$ for $z<0$.  The latter is the same as the particle
density profile in a right-moving TASEP at a $(2,1)$ junction, after making the
replacements $\rho\to 1-\rho$ and $z\to -z$.  Simulations show that the
density profile in this case is the mirror image of the density profile in
Fig.~\ref{rho21-1}(b)

An input density $\rho_{\scriptstyle L}<1$ in the $(1,2)$ junction geometry
corresponds to a $(2,1)$ junction with density $\rho=1$ for $z<0$ and
$\rho=1-\rho_{\scriptstyle L}$ for $z>0$; this correspondence is obvious
after making the replacements $\rho\to 1-\rho$ and $z\to -z$.  It is simpler
to describe the dynamics in terms of a right-moving TASEP at a $(2,1)$
junction with the initial condition of $\rho=1$ for $z<0$ and
$\rho=\rho_{\scriptstyle R}>0$ for $z>0$ and we do so in what follows.

For $t>0$, the density upstream from the junction again exhibits a pileup, in
which the scaled profile $f=1$ for $z<-1$, $f=\frac{1}{2}(1-z)$ for
$-1<z<-1/\sqrt{2}$, and finally $f=\rho_+$ for $-1/\sqrt{2}<z<0$.  For
$\rho_{\scriptstyle R}<\frac{1}{2}$, the incoming current equals its maximum
value of $\frac{1}{4}$.  This incoming current can be accommodated by a
rarefaction wave for $z>0$ that ends when the density decays to
$\rho_{\scriptstyle R}$.  On the other hand, for
$\rho_{\scriptstyle R}>\frac{1}{2}$, the outgoing current is density limited
and, therefore equal to
$j_{\scriptstyle R} =\rho_{\scriptstyle R}(1-\rho_{\scriptstyle R})$.  This
means that the pileup density $\rho_{\scriptstyle L}$ at $z=0^-$ is
determined by matching the currents at $z=0$.  This matching gives
$2\rho_{\scriptstyle L}(1-\rho_{\scriptstyle L})= j_{\scriptstyle R}$, or
$\rho_{\scriptstyle L}= \frac{1}{2}(1+\sqrt{1-2 j_{\scriptstyle R}})$, in
agreement with the density profile shown in Fig.~\ref{rho-1to2}(b).

\section{Open (2,1) Junction Geometry}
\label{sec:(2,1)-open}

We now study the $(2,1)$ junction with open boundary conditions with input
rate $\alpha$ and output rate $\beta$.  When the leftmost sites of the system
are empty, particles are inserted with rate $\alpha$; one could consider
distinct rates $\alpha_1$ and $\alpha_2$ for the two roads, but we limit
ourselves to the symmetric case of $\alpha_1=\alpha_2 = \alpha$.  Similarly,
when a particle reaches the rightmost site, it is extracted with rate
$\beta$.  These rates may take arbitrary positive values, but we limit
ourselves to the range $0<\alpha<1$ and $0<\beta<1$.  This restriction
corresponds to the system being coupled to reservoirs with particle density
$\alpha$ on the left and density $1-\beta$ on the right.

\begin{figure}[ht]
\includegraphics[width=0.35\textwidth]{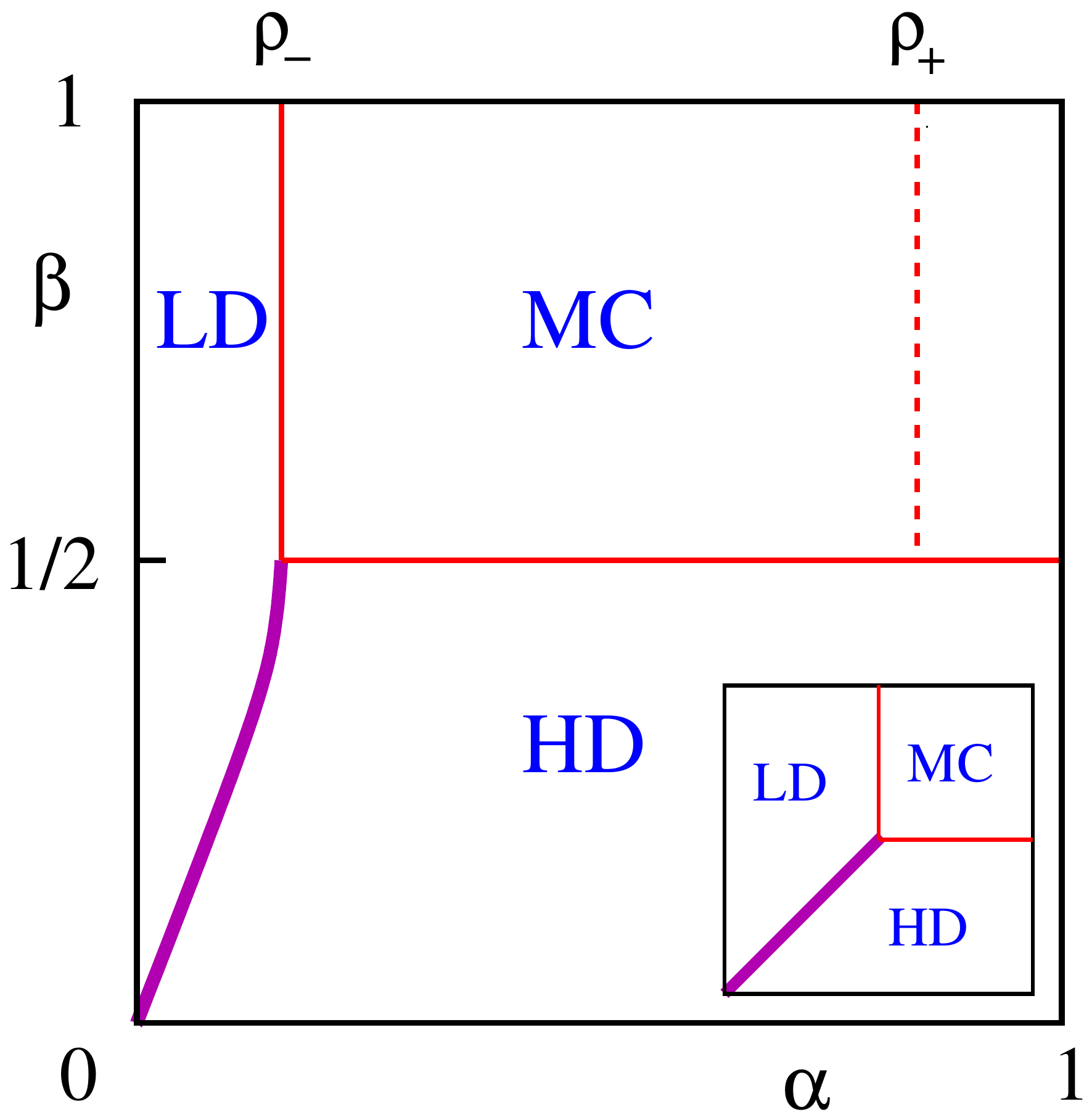}
\caption{Phase diagram for the open $(2,1)$ junction.  Inside the MC phase,
  $\alpha>\rho_-$ and $\beta>\frac{1}{2}$, the dashed vertical line at
  $\alpha=\rho_+$ separates a regime where the incoming road density is low,
  $\rho_{\scriptstyle L}=\rho_-$ when $\alpha<\rho_+$, from a high-density
  regime, $\rho_{\scriptstyle L}=\rho_+$ when $\alpha>\rho_+$.  The inset
  shows the corresponding phase diagram for the open single-road system.}
\label{Fig:PD}
\end{figure}

The behavior of this open system can be analyzed using the so-called domain
wall theory \cite{Krug91,KS2,Krug01,SA02,Sch03}; the basic predictions of
this theory agree with exact analyses (see \cite{BE07,Sch03} for reviews).
To put our results in context, it is helpful to first summarize the
properties of an open single-road TASEP.  Here, there are three phases
(Fig:~\ref{Fig:PD}): (i) a low-density (LD) phase, when $\alpha<\frac{1}{2}$
and $\alpha<\beta$; (ii) a high-density (HD) phase, when $\beta>\frac{1}{2}$
and $\alpha>\beta$; (iii) a maximal-current (MC) phase, when
$\alpha,\beta>\frac{1}{2}$.  For the $(2,1)$ junction, the same three phases
arise, but the locations of the phase boundaries are different than in a
single-road system.  The new feature for the $(2,1)$ junction geometry, which
already arose in the closed system, is that current conservation at the
junction leads to distinct densities just to the left and to the right of the
junction.

\bigskip\noindent \textbf{LD Phase:} $\alpha<\rho_-$ and
$\beta>\rho_{\scriptstyle R}(\alpha)$.  In the low-density phase, the exit
rate $\beta$ is relatively fast and the particle density is limited by the
rate at which particles enter the system.  Thus the density for $z<0$ is
$\rho_{\scriptstyle L}=\alpha$.  This statement holds as long as
$\alpha<\rho_-$, so that the current is less than $\frac{1}{4}$.  In this
case, the right-half of the system can support and transmit this incoming
current.  Using the current conservation statement \eqref{jlr} across the
junction, as well as $\rho_{\scriptstyle L}=\alpha$, we immediately obtain,
for the density in the right half of the system,
\begin{subequations}
\label{rho-R}
  \begin{equation}
  \rho_{\scriptstyle R}(\alpha) = \tfrac{1}{2}\left[1 - \sqrt{1-8\alpha(1-\alpha)}\right]\,.
\end{equation}
The current in this LD phase is $j(\alpha)=2\alpha(1-\alpha)$.

\bigskip\noindent \textbf{HD Phase:} $\alpha > \rho_{\scriptstyle L}(\beta)$
and $\beta<\frac{1}{2}$.  In the high-density phase, the exit rate $\beta$ is
relatively slow and the particle density is determined by the rate at which
particles enter the system.  In this HD phase, the density for $z>0$ is
$\rho_{\scriptstyle R}=\beta$.  We again invoke the current conservation
statement \eqref{jlr} across the junction and now solve for
$\rho_{\scriptstyle L}$ to give
\begin{equation}
\label{rho-L}
  \rho_{\scriptstyle L}(\beta) = \tfrac{1}{2}\left[1 - \sqrt{1-2\beta(1-\beta)}\right]\,.
\end{equation}
\end{subequations}
The current in this HD phase is $j(\beta)=\beta(1-\beta)$.

\begin{figure}[ht]
\includegraphics[width=0.35\textwidth]{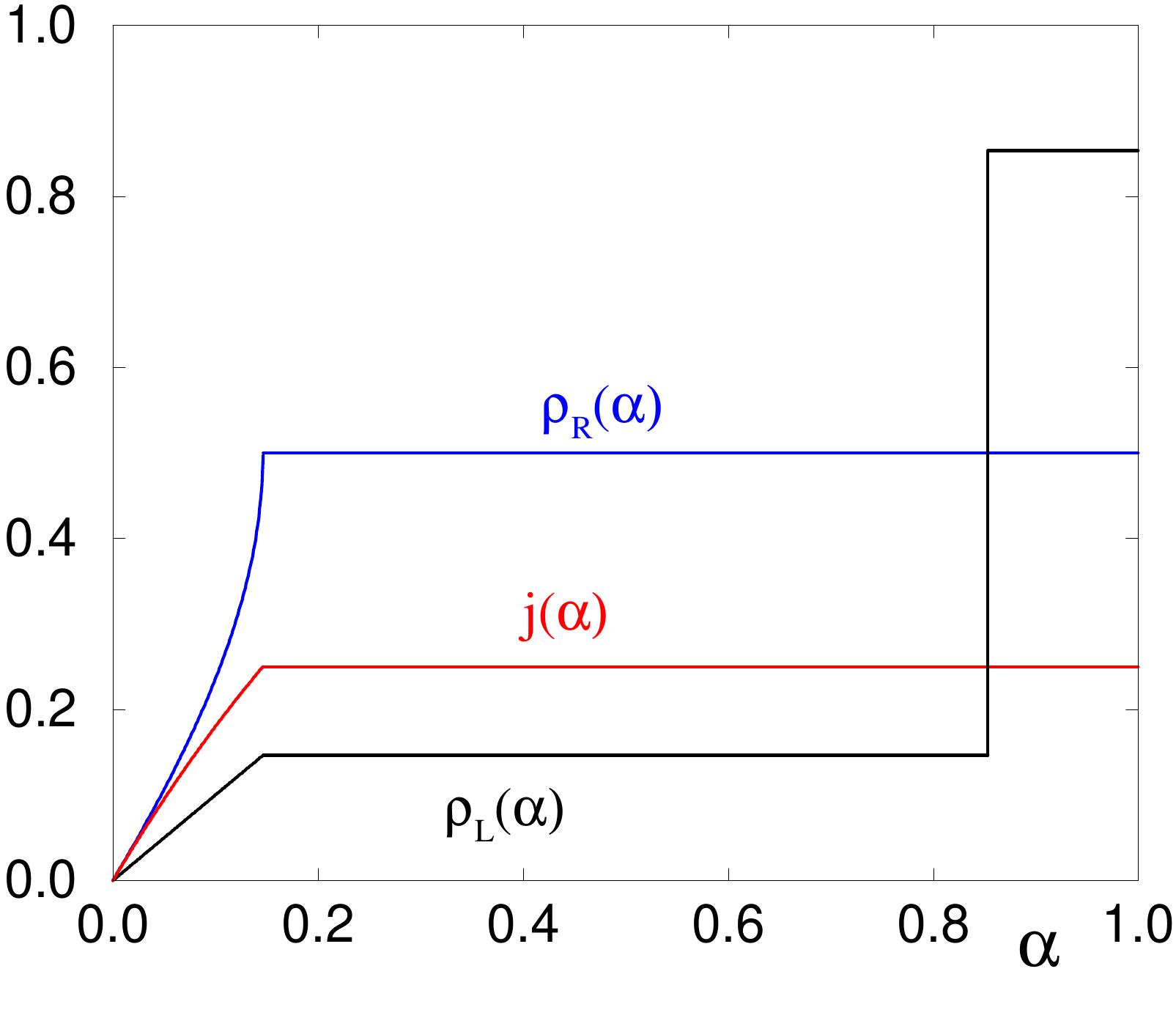}
\caption{The bulk densities
  $\rho_{\scriptstyle L}(\alpha), \rho_{\scriptstyle R}(\alpha)$, and the
  current $j(\alpha)$ when $\beta>\frac{1}{2}$. }
\label{fig:RJ}
\end{figure}

\bigskip\noindent
\textbf{MC phase:} $\alpha>\rho_-$ and $\beta>\frac{1}{2}$. The
density before the junction in the MC phase is
\begin{equation}
\label{L:MC}
 \rho_{\scriptstyle L} = 
 \begin{cases}
 \rho_-  & \text{when} ~~\rho_-<\alpha<\rho_+\\
 \rho_+ & \text{when} ~~\rho_+<\alpha
 \end{cases}
\end{equation}
The density in the two input roads can be either $\rho_-$ or $\rho_+$ to
ensure that the total incoming current is the maximal possible current in a
single road.  Interestingly, the density in the left half of the system
changes discontinuously when the input rate $\alpha=\rho_+$.  For both cases
the density in the right half of the system is
$\rho_{\scriptstyle R}=\frac{1}{2}$ and the current is $j=\frac{1}{4}$ (see
Fig.~\ref{fig:RJ}).

\bigskip\noindent 
\textbf{Coexistence line:} The co-existence line is
defined by the condition $\rho_{\scriptstyle R}(\alpha)=\beta$, with
$\rho_{\scriptstyle R}(\alpha)$ given by Eq.~\eqref{rho-R}, or equivalently
$\rho_{\scriptstyle L}(\beta)=\alpha$, with $\rho_{\scriptstyle L}(\alpha)$
given by Eq.~\eqref{rho-L}.  The additional conditions $\alpha<\rho_-$ and
$\beta<\frac{1}{2}$ must also hold.  This line (magenta in Fig.~\ref{Fig:PD})
separates the LD and HD phases.  In the case of the single-road TASEP, subtle
behaviors occur on the co-existence line $0<\alpha=\beta<\frac{1}{2}$. It is
known \cite{KS2} that the density profile is a stationary shock wave with
$\rho=\alpha$ near the left end and $\rho=1-\alpha$ near the right end.  The
location of the shock is a uniformly distributed random variable.  By
averaging over all possible locations of the shock for the open system on the
interval $-L<x<L$, the density is given by
\begin{equation}
\rho(x) = \frac{1}{2}+\left(\frac{1}{2} - \alpha\right)\frac{x}{L}\,.
\end{equation}

We anticipate a similar behavior on the co-existence line for the $(2,1)$
junction.  The density near the entrance is $\alpha$ and the density near the
exit is $1-\beta$. In the simplest situation when the shock is located at the
junction
\begin{equation}
\label{co-exist}
 \rho = 
\begin{cases}
\alpha  & \quad-L_1<x<0\\
1-\beta & ~\qquad 0<x<L_2\,.
 \end{cases}
\end{equation}
However, the distribution of the position of the shock front is unknown, and
the lengths $L_1$ and $L_2$ of the incoming and outgoing roads may play a
significant role.

\section{Discussion}

We introduced a simple extension of the totally asymmetric simple exclusion
process (TASEP), in which multiple roads meet at a junction.  We focused on
two simple geometries, namely the $(2,1)$ and the $(1,2)$ junctions, in which
either two roads merge into one, or one road splits into two.  We first
treated the density downstep initial condition, which normally leads to a
rarefaction wave.  For the $(2,1)$ junction geometry, we found much richer
behavior in which there can be a particle pileup upstream from the junction.
Additionally a shock wave can arise for a suitable range of initial
densities.  These phenomena qualitatively resemble what occurs in real
traffic that approaches a constriction on a highway.

We also investigated non-equilibrium steady states in the $(2,1)$ junction
with open boundaries in which particles are continuously fed in at the left
end and removed from the right end. We analyzed this system using domain-wall
theory~\cite{Krug91,KS2,Krug01,SA02,Sch03}, which is known to correctly
predict the phase diagram for the TASEP and more complicated lattice gases.
It would be desirable to study junctions with open boundaries using exact
approaches.  One possibility is to attempt to extend the matrix product
approach~\cite{DEHP} to the junction geometry.  If the matrix product
formulation can be extended to the junction geometry, then it should be
feasible to provide detailed insights into the spatial structure of the
non-equilibrium steady states.  For example, the matrix product approach
should be able to give the behavior of the density profile in the boundary
layers near the left and right ends of the system, and in the inner layer
near the junction.  In the single-lane TASEP, these boundary layers exhibit
qualitative changes within the low-density and high-density phases: the LD
phase may be subdivided into LD-I and LD-II, and similarly for the HD phase.
A more accurate description of the open $(2,1)$ junction phase diagram may
perhaps be richer still due to the potentially different behaviors in the
inner layer on either side of the junction.

Apart from generalizations to the $(m,n)$ junction geometry and to models
with different hopping rates in the incoming and outgoing roads (which could
be viewed as different speed limits in the two types of roads), one can probe
the influence of the coupling between the parallel lanes in the bulk, in
addition to the local interaction at the junction site. Multilane models can
exhibit rich behaviors even without junctions (see, e.g.,
\cite{PS04,RJ07,EKST,AES15}).  Junction-like geometries have been recently
investigated in modeling pedestrian traffic~\cite{Apper11,Apper13}; however,
the movement rules in these pedestrian movement models were significantly
different from TASEP dynamics.

It would be also interesting to study the TASEP on more complicated graphs
with vertices mimicking junctions. One amusing example is the TASEP on a
figure-eight geometry, in which a particle can pass through the junction of
the figure eight only when it is clear.  This geometry is inspired by the
infamous automobile races on the Islip Figure-Eight Speedway~\cite{islip}
that were held between 1962 and 1984.  The course is in the shape of a figure
eight, with a collision point where the two loops of the figure eight meet.
In this figure-eight geometry, we anticipate large collision-induced temporal
fluctuations in the current passing through the junction.

Finally we emphasize that our analysis of junction geometries relied on
hydrodynamic techniques, which yield only average characteristics.
Fluctuations in the TASEP have attracted considerable interest.  For example,
for the density downstep initial condition, the total number of particles
$N(t)$ that flow to the initially empty half-line by time $t$ is a random
quantity whose fluctuations scale as $t^{1/3}$; that is,
\begin{equation}
\label{N:fluct}
N(t) = \tfrac{1}{4}\,t + t^{1/3}\xi\,.
\end{equation}
The distribution $P(\xi)$ of the random variable $\xi$ was established by
Johansson~\cite{Johan}.  Remarkably, the same and related Tracy-Widom
distributions were derived earlier in the context of random
matrices~\cite{TW94}, and they arise in a wider range of problems
(see~\cite{spohn,Ivan_rev} and references therein).  For the $(2,1)$ junction
we anticipate the same functional form \eqref{N:fluct}, but the distribution
of the corresponding random variable is unknown.  For the $(1,2)$ junction,
we expect that the total numbers of particles in each of the the two roads
are
\begin{equation}
\label{N12:fluct}
N_1(t) = \tfrac{1}{8}\,t + t^{1/3}\eta_1, \quad N_2(t) = \tfrac{1}{8}\,t + t^{1/3}\eta_2\,.
\end{equation}
The random variables $\eta_1$ and $\eta_2$ are correlated, and computing the
joint distribution $P(\eta_1,\eta_2)$ appears to be challenging.

\begin{acknowledgments}
  KZ gratefully acknowledges the support of an Ariel Scholarship through St.\
  John's College, Santa Fe.  He also thanks the Research Experience for
  Undergraduates program at the Santa Fe Institute that was funded by NSF
  grant OAC-1757923.  PLK thanks the hospitality of the Santa Fe Institute
  where this work was completed.  SR gratefully acknowledges financial
  support from NSF grant DMR-1608211.  We also thank Joachim Krug, Joel
  Lebowitz and Kirone Mallick for helpful suggestions and correspondence.
\end{acknowledgments}

\end{document}